\documentclass[twocolumn]{revtex4}

\usepackage{bm}

\newcommand{\ket}[1]{|#1\rangle}
\newcommand{\ex}[1]{\langle#1\rangle}
\renewcommand{\d}{{\rm d}}
\newcommand{\e}{{\rm e}}
\newcommand{\sinc}{{\rm sinc}}

\begin{document}

\title{Lower limit on decoherence introduced
by entangling two spatially-separated qubits}

\author{A.J.~Fisher}
\email{andrew.fisher@ucl.ac.uk}
\homepage{http://www.cmmp.ucl.ac.uk/~ajf}
\affiliation{Department of Physics and Astronomy, University College
London,\\
Gower Street, London WC1E 6BT, U.K.}

\begin{abstract}
It is shown that a generalization of the fluctuation-dissipation
theorem places an upper bound on the figure of merit for any
quantum gate designed to entangle spatially-separated qubits.  The
bound depends solely on the spectral properties of the
environment. The bound applies even to systems performing a
quantum computation within a decoherence-free subspace, but might
be optimized by the use of non-equilibrium squeezed states of the
environment, or by using an environment system constructed to have
response confined to certain frequencies.
\end{abstract}

 \maketitle

Many possible systems to realise the ultimate goal of quantum
computation \cite{nielsenbook} have been proposed in recent years.
In choosing systems, one has to trade off two competing
requirements: one would like to be able to manipulate the state of
the quantum bits (qubits) rapidly, in order to be able to perform
the computation as quickly as possible.  On the other hand, it is
important to avoid decoherence. Typically, qubits which interact
relatively strongly with their surroundings can be manipulated
more rapidly than those that do not, but also decohere more
rapidly; qubits based on excitons in quantum dots
\cite{bonadeo98,quiroga99}, or on nuclear spins
\cite{cory97,jones00}, provide examples of the fast and slow
extremes respectively. In many cases the important quantity is the
ratio of the decoherence time to the characteristic manipulation
time, often referred to as the `figure of merit'.

In this paper we make precise the connection between the rate at
which two spatially separated qubits can be entangled, and the
minimum amount of decoherence that is necessary to accomplish
this, thereby placing an upper bound on the figure of merit.  This
connection is inevitable because spatially separated systems
cannot be coupled directly; instead, any coupling between them
must be mediated by some third system (the environment). For two
atoms in a vacuum the environment is the electromagnetic field;
for qubits in a solid it might instead be the phonons or
conduction electrons. The capacity of the environment to change
state in response to one qubit, and the subsequent effect of this
change on the other qubit, is what produces the coupling and leads
to entanglement of the qubits; yet it also leads to the qubits and
the environment becoming entangled, and hence to the decoherence
of the qubits. (There may also be additional sources of
decoherence, not related to the coupling.) It is worth noting from
the outset that in order to make this connection quantitative, it
is essential to go beyond Markovian formulations of decoherence,
such as that of Lindblad \cite{lindblad76}, because they take no
account of the history of the environment. Therefore, they cannot
represent the `feedback' of the environment on the system that is
the essential feature of the communication between the qubits.

Consider two systems, $A$ and $B$, that encode the quantum
information.  Suppose they are not coupled directly, but only
indirectly via their mutual coupling to an environment $E$:
\begin{equation}\label{generalh}
\hat H=\hat{H}_A+\hat{H}_B+\hat{H}_E+\lambda(\hat{H}_{EA}+\hat{H}_{EB}).
\end{equation}
We assume the Hamiltonian is time-independent; our argument can be
generalized to the case where $A$ and $B$ are driven by a
time-dependent external (classical) field.  Let us write the
coupling Hamiltonian for the $A$ system to the environment as
\begin{equation}\label{perturbation}
\hat{H}_{EA}=\hat{V_A}\otimes\hat{O_A},
\end{equation}
where $\hat{V}_A$ is a Hermitian system $A$ operator and $O_A$ is
a system $E$ operator, and correspondingly for system $B$.  (The
most general form for the coupling is a sum of different terms of
this type, but our discussion may be easily generalized to that
case.)

Suppose the initial ($t=0$) density matrix for the system consists
of a direct product of some (possibly mixed) state of $A$ and $B$
with a (possibly mixed) state of the environment $E$.  We assume
that this state of $E$ is a thermal equilibrium distribution, so
we can write
\begin{equation}
\hat{\rho}(0)=\rho_{AB}(0)\otimes{\exp(-\beta\hat{H}_E)\over Z_E},
\end{equation}
where $\beta=1/k_BT$ and $Z_E$ is the environment's partition
function.  Then, treating the coupling
$\lambda(\hat{H}_{EA}+\hat{H}_{EB})$ as the perturbation and
working in the interaction representation, the reduced density
matrix $\rho_{AB}$ of the $AB$ system evolves as
\begin{equation}
\hat{\rho}_{AB}(t)=\hat{\rho}_{AB}(0)+\Delta\hat{\rho}_{AB}^{(1)}(t)
+\Delta\hat{\rho}_{AB}^{(2)}(t)+ {\rm O}(\lambda^3).
\end{equation}
The first-order term $\Delta\hat{\rho}_{AB}^{(1)}$ is given by
\begin{equation}\label{firstorder}
\Delta\hat{\rho}_{AB}^{(1)}(t)=-\mathrm{i}\lambda\int_0^{t}\d
t'\biggl[\ex{\hat{O}_A}
\hat{V}_A(t')\hat{\rho}(0)+\ex{\hat{O}_B}\hat{V}_B(t')\hat{\rho}(0)
\biggr]\nonumber\\
+\hbox{h.c.},
\end{equation}
where the angle brackets $\ex{\ldots}$ correspond to the
equilibrium average in the isolated environment. We keep $\hbar=1$
throughout. Equation (\ref{firstorder}) simply corresponds to the
evolution of the $AB$ system in the static `applied field' of the
environment. This evolution could equally well have been absorbed
into that generated by $\hat{H}_A$ and $\hat{H}_B$ and leads
neither to entanglement between $A$ and $B$ nor to decoherence
(provided that any external fields entering $\hat{H}_A$ and
$\hat{H}_B$ are sufficiently `classical' \cite{vanenk01}).  We
will henceforth assume that $\ex{\hat{O}_A}=\ex{\hat{O}_B}=0$, and
neglect this ${\rm O}(\lambda)$ term.

To second order, we obtain
\begin{eqnarray}\label{generalsecondorder}
&&\Delta\rho^{(2)}(t)=\sum_{ab}\lambda^2\biggl[\int_0^t\d
t'\int_0^t\d t''\nonumber\\
&&\quad\hat{V}_a(t')
\hat{\rho}_{AB}(0)\hat{V}_b(t'')C_{ba}(t''-t')\nonumber\\
&&-\int_0^t\d t'\int_{t'}^{t}\d t''\nonumber\\
&&\quad\biggl\{\hat{V}_a(t'')V_b(t')\hat{\rho}_{AB}(0)C_{ab}(t''-t')
+\hbox{h.c.}\biggl\}\biggr].
\end{eqnarray}
The indices $\{a,b\}$ both run over $A$ and $B$, and the
environmental correlation functions $C_{ab}(t)$ are defined as
$C_{ab}(t)\equiv\ex{\hat{O}_a(t)\hat{O}_b(0)}=\ex{\e^{\mathrm{i}
\hat{H}_Et}\hat{O}_a\e^{-\mathrm{i} \hat{H}_Et}\hat{O}_b}$.  The
contribution to $\hat{\rho}_{AB}$ in equation
(\ref{generalsecondorder}) is much more interesting, as the cross
terms involving $A$ and $B$ can entangle the $A$ and $B$ systems.
However, these second-order contributions also inevitably decohere
the $AB$ system, as we shall see.

We now consider a specific example in which $A$ and $B$ are single
spin-1/2 qubits.  We suppose that each spin experiences an
associated classical magnetic field:
\begin{equation}
\hat{H}_A=-\gamma {\bf B}_A\cdot\hat{\sigma}^A,\qquad
\hat{H}_B=-\gamma {\bf B}_B\cdot\hat{\sigma}^B.
\end{equation}
We define $\omega_0$ to be the single-spin-flip energy in this
external field: $\omega_0\equiv 2\gamma|{\bf B}|$.  Initially we shall
suppose that the applied field is in the positive $z$-direction.  We further
suppose that the effect of the interaction with the environment is to
produce spin-flips; i.e., we choose
\begin{equation}
\hat{V}_A=\hat{\sigma}_x^A;\qquad \hat{V}_B=\hat{\sigma}_x^B.
\end{equation}
Note with this choice the coupling to the environment is
perpendicular to the applied field; this will lead to a $T_1$-type
relaxation of the spins.  Other choices of coupling (including
more realistic ones and those that would lead to $T_2$-type
relaxation) are possible, but do not change the essential features
of the result. However, we leave the form of the environment
itself completely general at this stage. Re-writing the
perturbation in terms of the raising and lowering operators
($\hat{\sigma}_x(t)=\hat{\sigma}_+\e^{-\mathrm{i}\omega_0t}+\hat{\sigma}_-\e^{\mathrm{i}\omega_0t}$)
and introducing two new indices $s$ and $s'$ which run over + and
-, we can perform the time integrals to obtain
\begin{eqnarray}\label{introducephipsi}
\Delta\hat{\rho}_{AB}^{(2)}(t)&=&\sum_{ab}\sum_{s,s'}\lambda^2
\int_{-\infty}^{\infty}{\d\omega\over 2\pi}J_{ab}(\omega)\nonumber\\
&&\times\biggl[\hat{\sigma}_{s}^{b}
\hat{\rho}_{AB}(0)\hat{\sigma}_{s'}^a\psi_{s,s'}(t,\omega)\nonumber\\
&&-\biggl\{\hat{\sigma}_{s'}^a\hat{\sigma}_{s}^b\rho_{AB}(0)
\phi_{s,s'}(t,\omega)+\hbox{h.c.}\biggl\}\biggr].
\end{eqnarray}
Here the power spectrum $J_{ab}(\omega)$ is the Fourier transform
of the corresponding correlation function
\begin{equation}
J_{ab}(\omega)\equiv\int_{-\infty}^{\infty}\d t
C_{ab}(t)\e^{\mathrm{i}\omega t},
\end{equation}
and the quantities $\psi$ and $\phi$ arise from the time integrals:
\begin{eqnarray}
\psi_{ss'}(t,\omega)&\equiv&\int_0^t\d t'\int_0^t\d t''\e^{[-\mathrm{i}\omega_0(st'+s't'')
-\mathrm{i}\omega(t''-t')]};\nonumber\\
\phi_{ss'}(t,\omega)&\equiv&\int_0^t\d t'\int_{t'}^t\d t''\e^{[-\mathrm{i}\omega_0(st'+s't'')
-\mathrm{i}\omega(t''-t')]}.
\end{eqnarray}

{\it Continuous spectrum.}  We first of all treat the case of a
continuous spectrum $J_{ab}(\omega)$ (i.e., one in which $J$ does
not vary significantly over a frequency range $\sim t^{-1}$). For
times that are long in comparison with the characteristic inverse
frequencies of the environment (but still short compared with the
characteristic motions of the qubits, so that the second-order
expansion in $\lambda$ is adequate), the contributions from $\phi$
and $\psi$ with $s=s'$ are vanishingly small and we may
approximate
\begin{eqnarray}
\psi_{s,-s}(t,\omega)&\approx& {-\mathrm{i} t\over(\omega-s\omega_0)}+\pi
t\delta(\omega-s\omega_0);\nonumber\\
\psi_{s,-s}(t,\omega)&\approx& 2\pi t\delta(\omega-s\omega_0).
\end{eqnarray}
The second-order result then simplifies to
\begin{equation}
\Delta\hat{\rho}_{AB}^{(2)}(t)=\Delta\rho_{+-}+\Delta\rho_{-+},
\end{equation}
where the subscripts refer to the values of $s,s'$ giving rise to
the contributions, and
\begin{eqnarray}
\Delta\rho_{s,-s}&=&\lambda^2t\sum_{ab}\biggl[\bigl\{\hat{\sigma}_s^b\hat{\rho}(0)\hat{\sigma}_{-s}^a
-{1\over2}\hat{\sigma}_{-s}^a\hat{\sigma}_s^b\hat{\rho}(0)\nonumber\\
&&-{1\over2}\hat{\rho}(0)\hat{\sigma}_{-s}^a\hat{\sigma}_{s}^b\bigr\}J_{ab}(s\omega_0)\nonumber\\
&&-\mathrm{i}\int{\d\omega\over2\pi}{J_{ab}(\omega)\over\omega-s\omega_0}
\bigl\{\hat{\sigma}_{-s}^a\hat{\sigma}_{s}^b\hat{\rho}(0)-\hat{\rho}(0)\hat{\sigma}_{-s}^{a}\hat{\sigma}_{s}^b\bigr\}\nonumber\\
\end{eqnarray}
This can be rewritten in the effective Lindblad form \cite{lindblad76}
\begin{equation}\label{lindbladeq}
\Delta\hat{\rho}_{AB}=t\biggl\{-\mathrm{i} [\hat{H}_{\rm
eff},\hat{\rho}_{AB}] +\sum_\mu\biggl(\hat{L}_\mu
\hat{\rho}_{AB}\hat{L}_\mu^\dag -{1\over2}\{\hat{L}_\mu^\dag
\hat{L}_\mu,\rho_{AB}\}\biggr)\biggr\},
\end{equation}
where the components of the effective Hamiltonian $\hat{H}_{\rm
eff}=\hat{H}_{-+}+\hat{H}_{+-}$ are
\begin{eqnarray}\label{Heff}
\hat{H}_{+-}&=&\lambda^2\sum_{ab}\int_{-\infty}^\infty{\d\omega\over2\pi}
{J_{ab}(\omega)\over(\omega-\omega_0)}\hat{\sigma}_-^a\hat{\sigma}_+^b
\nonumber\\
\hat{H}_{-+}&=&\lambda^2\sum_{ab}\int_{-\infty}^\infty{\d\omega\over2\pi}
J_{ab}(\omega){1\over(\omega+\omega_0)}\hat{\sigma}_+^a\hat{\sigma}_-^b\nonumber\\
&=&-\lambda^2\sum_{ab}\int_{-\infty}^\infty{\d\omega\over2\pi}{J_{ab}(\omega)\over(\omega-\omega_0)}
\e^{-\beta\omega}\hat{\sigma}_+^b\hat{\sigma}_-^a.
\end{eqnarray}
Provided the $J_{AB}\ne0$ (i.e. provided that the fluctuations
experienced by $A$ and $B$ are correlated), $\hat{H}_{\rm eff}$
can produce entanglement between $A$ and $B$, as desired.  But
there are also four Lindblad operators $L_\mu$, two for positive
frequencies (linear combinations of $\hat{\sigma}_+^{a,b}$,
corresponding to correlated spin flips in the direction of the
field) and two for negative frequencies (linear combinations of
$\hat{\sigma}_-^{a,b}$, correlated spin flips opposite to the
field). The combinations are those that diagonalize the quadratic
forms
\begin{eqnarray}\label{quadform}
\sum_\mu L_{\mu,+}^\dag
L_{\mu,+}&=&\lambda^2\sum_{ab}J_{ab}(\omega_0)\hat{\sigma}_-^{a}\hat{\sigma}_+^{b}\nonumber\\
\sum_\mu L_{\mu,-}^\dag
L_{\mu,-}&=&\lambda^2\sum_{ab}J_{ab}(-\omega_0)\hat{\sigma}_+^{a}\hat{\sigma}_-^{b}\nonumber\\
&=&\e^{-\beta\omega_0}\lambda^2\sum_{ab}J_{ab}(\omega_0)\hat{\sigma}_+^{b}\hat{\sigma}_-^{a}
\end{eqnarray}
If the environments of $A$ and $B$ are identical, symmetry
requires that the appropriate combinations are proportional to
$\hat{\sigma}_{\pm}^A\pm\hat{\sigma}_{\pm}^B$ .  One can now see
clearly the main point of this paper: in contrast to the Markovian
approximation \cite{lindblad76}, the effective Hamiltonian
(generating the coherent evolution of the two qubits) and the
Lindblad operators (determining the incoherent decay) are not
independent, because they are determined by the {\em same
correlation functions $J_{ab}(\omega)$ of the environment}. It is
therefore in general impossible to have one without the other. The
ratio between the two, and hence the maximum obtainable figure of
merit for the gate, is independent of the coupling strength
$\lambda$ to the environment; instead, it is determined by the
spectral characteristics of the environmental fluctuations.

These results are intimately connected to the
fluctuation-dissipation theorem \cite{kubo57}.  The environment
develops a `response' to the applied `field' from the first qubit,
and it is this response that then drives the second qubit into a
state entangled with the first.  The change in the expectation
value $\ex{\hat{O}_B}$ when a classical perturbation $\hat{O}_A$
is applied to the environment is determined the
frequency-dependent susceptibility
\begin{equation}\label{chidef}
\chi_{BA}(\omega)=\int_{-\infty}^\infty\d\omega'
{(1-\e^{-\beta\omega'})J_{BA}(\omega')\over\omega-\omega'}.
\end{equation}
Note that, since the applied field is classical, $\chi$ contains
both a positive-frequency and a negative-frequency part, the
latter being suppressed by a factor $\e^{-\beta\omega}$. The real
and imaginary parts of $\chi$ are connected by dispersion
relations, and hence the dissipative (imaginary) part of $\chi$ is
directly connected to the fluctuation spectrum by
\begin{equation}\label{fdt}
\Im(\chi_{AB}(\omega))+\Im(\chi_{BA}(\omega))=[J_{AB}(\omega)n(\omega)+J_{BA}(-\omega)n(-\omega)].
\end{equation}
Here $n(\omega)\equiv[\exp(\beta\omega)-1]^{-1}$ is the Bose
occupation number at frequency $\omega$.

In our case, we have to distinguish between positive-frequency
(environment transiently absorbs energy from system $A$ or $B$)
and negative-frequency (environment gives out energy) processes.
Both contribute to the effective Hamiltonian (\ref{Heff}), but
each corresponds to a different type of decoherence.  There are
therefore two separate relations that link the effective
Hamiltonian when the system is operated at some given frequency
$\omega$ to the positive- and negative-frequency incoherent terms
at all other frequencies $\omega'$:
\begin{eqnarray}
\hat{H}_{s,-s}(\omega)&=&\int_{-\infty}^{\infty}{\d\omega'\over
2\pi}{\bigl[\sum_\mu L_{\mu,s}^\dag(\omega')
L_{\mu,s}(\omega')\bigr]\over\omega'-\omega}; \quad s\in\{+,-\}.\nonumber\\
\end{eqnarray}
A bound on the figure of merit for any particular gate can now be
obtained by comparing the time needed for the gate operation to be
generated by $\hat{H}_{\rm eff}$ with the decoherence times
governed by the Lindblad operators.

{\it Discrete spectrum.}  The opposite limit is that in which the
power spectra of the correlation functions vary much more rapidly
with frequency than do the functions $\psi_{ss'}(t,\omega)$ and
$\phi_{ss'}(t,\omega)$.  In particular, if the frequency response
of the environment is strictly confined to discrete frequencies
$\Omega_n$, we can write
\begin{eqnarray}
J_{ab}(\omega)=\sum_nJ_{ab,n}\delta(\omega-\Omega_n),
\end{eqnarray}
but must keep the full spectral form of the functions $\phi$:
\begin{eqnarray}
\phi_{s,-s}(t,\omega)&=&{2\over(\omega-s\omega_0)^2}\sin^2[(\omega-s\omega_0)t]\nonumber\\
&&+\mathrm{i}\biggl[{-t\over\omega-s\omega_0}+{1\over(\omega-s\omega_0)^2}
\sin\biggl({(\omega-s\omega_0)t\over 2}\biggr)\biggr];\nonumber\\
\phi_{s,s}(t,\omega)&=&{-1\over
2\omega_0}\biggl({1\over\omega-s\omega_0}[\e^{-2\mathrm{i} s\omega_0t}-
\e^{-\mathrm{i}(s\omega_0+\omega)t}]\nonumber\\
&&-{1\over\omega+s\omega_0}[\e^{-\mathrm{i}(s\omega_0+\omega)t}-1]\biggr)\nonumber\\
\psi_{s,-s}(t)&=&{4\over(\omega-s\omega_0)^2}\sin^2\left[{(\omega-s\omega_0)t\over2}\right];\nonumber\\
\psi_{s,s}(t,\omega)&=&{-\mathrm{i}
s\over2\omega_0}\left\{{1\over\omega+s\omega_0}-
{1\over\omega-s\omega_0}\right\}\nonumber\\
&&\times(\e^{-2\mathrm{i}
s\omega_0t}-\e^{\mathrm{i}(\omega-s\omega_0)t}-\e^{-\mathrm{i}(s\omega_0+\omega)t}+1).
\end{eqnarray}
Evaluating equation~(\ref{introducephipsi}) for this case, we find
\begin{eqnarray}
\Delta\hat{\rho}_{AB}^{(2)}(t)&=&\sum_{ab}\sum_{s,s'}\lambda^2
\sum_n{1\over 2\pi}J_{ab,n}\nonumber\\
&&\quad\biggl[\hat{\sigma}_{s}^{b}
\hat{\rho}_{AB}(0)\hat{\sigma}_{s'}^a\psi_{s,s'}(t,\Omega_n)\nonumber\\
&&-\biggl\{\hat{\sigma}_{s'}^a\hat{\sigma}_{s}^b\hat{\rho}_{AB}(0)
\phi_{s,s'}(t,\Omega_n)+\hbox{h.c.}\biggl\}\biggr].
\end{eqnarray}
This also can be written in the form of
equation~(\ref{lindbladeq}), albeit with a time-dependent
effective Hamiltonian
\begin{equation}
\mathrm{i} t\hat{H}_{\rm eff}=-{1\over2}\sum_n\sum_{ab}\sum_{ss'}
{J_{ab,n}\over 2\pi}
(\phi_{ss'}-\phi_{-s',-s}^*)\hat{\sigma}_{s'}^{a}\hat{\sigma}_s^{b};
\end{equation}
and time-dependent effective Lindblad operators:
\begin{eqnarray}
&&t\sum_\mu\bigl[\hat{L}_\mu\hat{\rho}_{AB}\hat{L}_\mu^\dag-{1\over2}\{\hat{L}_\mu^\dag\hat{L}_\mu,\hat{\rho}_{AB}\}\bigr]\hfill\nonumber\\
&&=\sum_n\hfill\sum_{ab}\sum_{ss'}J_{ab}\psi_{ss'}\bigl[
\hat{\sigma}_s^{b}\hat{\rho}_{AB}\hat{\sigma}_{s'}^{a}-{1\over2}\{\hat{\sigma}_{s'}^{a}\hat{\sigma}_s^{b},\hat{\rho}_{AB}\}\bigr].
\end{eqnarray}
Note we have used the result
$\psi_{ss'}=\phi_{ss'}+\phi_{-s',-s}^*$. Once again the
Hamiltonian and the Lindblads depend on the same correlation
functions, although now two possible strategies are apparent that
may suppress the relative magnitude of the Lindblads. One is to
ensure that the environment contains no frequency $\Omega_n$
nearly resonant with the transition frequency $\omega_0$; the
dominant terms in the Lindblads then scale as
$t^2\sinc^2(\Delta\omega t)$, where $\Delta\omega$ is the
frequency interval between the operating frequency $\omega_0$ and
the nearest environment frequency $\Omega_n$. A second strategy is
to ensure that the gate's operating timescale coincides with the
zeros of $\psi_{ss'}(t,\omega)$, so that there is no decoherence
following one complete operation cycle (although there may be some
at intermediate times). This is only likely to be possible if a
single frequency $\Omega_n$ dominates the response. Note that this
second strategy guarantees a zero contribution to the decoherence
from frequency $\Omega_n$, regardless of the magnitude of the
fluctuations $J_{ab,n}$.  This is how a `warm' ion vibrational
mode is exploited to produce entanglement in an ion trap
\cite{molmer99,sackett00}; our argument shows that this idea is
much more general.

Our results may be considered a generalization of the
`fluctuation-dissipation-entanglement theorem' proposed by Sidles
{\it et al.} \cite{sidles00}; these authors, however, considered
just the interaction of a single spin with a harmonic bath.  The
present results confirm their expectation that such connections
could be found for arbitrary environments. We also note that
particular cases of the fundamental connection between coherent
and incoherent evolution have been noticed before, including the
Korringa law connecting the relaxation rate to the Knight shift
for NMR in metals \cite{slichterbook}, single-qubit manipulations
in quantum optics \cite{vanenk01}, and limits placed by
spontaneous emission on quantum manipulations in ion traps
\cite{plenio97}.

Unfortunately, the elegant device of the Decoherence Free Subspace
(DFS) \cite{lidar98} is of no use in avoiding the type of
irreducible decoherence we have described.  Using a DFS, we would
work entirely with states $\ket{\psi}$ of the two qubits for which
\begin{equation}
\hat{V}_A\ket{\psi}=\hat{V}_B\ket{\psi}=0.
\end{equation}
However, there would then be no way to entangle the two qubits by
the coherent part of the evolution, since $\hat{H}_{\rm
eff}\ket{\psi}=0$. Therefore, the DFS is only useful so long as
the dominant source of decoherence is {\it different\/} from the
environmental response producing the inter-qubit coupling; the DFS
can reduce the decoherence from other sources, but not eliminate
the irreducible component described here.

Our results suggest several ways to minimize the unavoidable
decoherence. Two, described above, involve engineering the
frequency spectrum of the environment or carefully choosing the
operation time of the gates. Another route might be by preparing
special initial `squeezed' states of the environment, in which the
fluctuations in those observables that couple to the qubits are
reduced below their equilibrium values (even, in principle, below
their zero-temperature values), at the expense of larger
fluctuations in other (conjugate) variables that do not couple to
the qubits.  Application of these ideas to bound the figures of
merit for various specific types of inter-qubit coupling will be
described in a separate publication.    However, it should also be
noted that the present results come from a weak-coupling expansion
in the interaction between the qubits and the environment; it is
entirely possible that very strong interactions in certain systems
might lead to a different balance between entanglement and
decoherence.

The author would like to thank Marshall Stoneham, Joe Gittings,
Thornton Greenland, Steve Cox, Tim Spiller, Bill Munro, Rasmus
Hansen, David Pettifor and Martin Plenio for discussions and
suggestions.  This work was partially supported by the EPSRC under
GR/M67865, and partly by the IRC in Nanotechnology.

\bibliography{qc}

\end{document}